\documentclass[a4paper,fleqn]{cas-sc}

\usepackage[numbers,sort&compress]{natbib}
\usepackage{lipsum}
\usepackage{subcaption}
\usepackage{amsmath}
\usepackage{amsfonts}
\usepackage{upgreek}
\usepackage{textcomp}

\def\tsc#1{\csdef{#1}{\textsc{\lowercase{#1}}\xspace}}
\tsc{WGM}
\tsc{QE}

\begin{document}

\let\WriteBookmarks\relax
\def\floatpagepagefraction{1}
\def\textpagefraction{.001}


\shorttitle{Piezoelectric composite cements}    

\shortauthors{Triana-Camacho \textit{et al.}} 

\title[mode = title]{Piezoelectric composite cements: Towards the development of self-powered and self-diagnostic materials}  



\author[label1]{Daniel A. Triana-Camacho}[orcid=0000-0001-6852-6277]
\ead{dantrica@saber.uis.edu.co}
\author[label1]{Jorge H. Quintero-Orozco}[orcid=0000-0002-9394-4515]
\ead{jhquinte@uis.edu.co}
\author[label2]{Enrique Mejía-Ospino}
\author[label3]{Germán Castillo-López}
\author[label4]{Enrique Garc\'{i}a-Mac\'{i}as\corref{cor1}}[orcid=0000-0001-5557-144X]
\ead{enriquegm@ugr.es}

\address[label1]{Ciencia de Materiales Biológicos y Semiconductores (CIMBIOS), Escuela de F\'{i}sica, Universidad Industrial de Santander, Cra 27 Calle 9, Bucaramanga, Colombia.}
\address[label2]{Laboratorio de Espectroscopía Atómica y Molecular, Escuela de Química, Universidad Industrial de Santander, Bucaramanga, Colombia.}
\address[label3]{Escuela de Ingenierías Industriales, Departamento de Ingeniería Civil de Materiales y Fabricación, Calle Dr. Ortiz Ramos s/n, 29071, Málaga, Spain.}
\address[label4]{Department of Structural Mechanics and Hydraulic Engineering, University of Granada, Av. Fuentenueva sn, 18002 Granada, Spain.}
\cortext[cor1]{Corresponding author. Department of Structural Mechanics and Hydraulic Engineering, University of Granada, Av. Fuentenueva sn, 18002 Granada, Spain. Phone: +39 958241000 ext. 20668 }


\begin{abstract}
Piezoresistivity is the most commonly used sensing principle in cement-based smart composites for strain-monitoring applications. Nonetheless, the need for external electric power to conduct electrical resistivity measurements restricts the scalability of this technology, especially when implemented in remote structures. To address this issue, this manuscript thoroughly analyzes the piezoelectric properties of cement composites doped with reduced graphene oxide (rGO) and evaluates their potential as self-powered strain sensors. To do so, a comprehensive methodology involving voltammetry measurements, open circuit potential determination, and uniaxial compression testing is developed to determine the piezoelectric coefficients of charge $d_{33}$ and voltage $g_{33}$. Furthermore, a novel circuital model for signal processing of the electromechanical response is developed and experimentally validated in terms of time series of output voltage, resistance, and the generated electric power. The developed methodology is applied to laboratory samples manufactured following two different filler dispersion methods. The presented results evidence that samples prepared by ultrasonic cleaner dispersion achieve optimal properties, with a piezoelectric charge coefficient of 1122.28$\mathrm{\pm}$246.67 pC/N, about 47 times greater than previously reported composites in the literature. Unlike piezoresistive cement-based composites, a remarkable nonlinear correlation between the fractional change in the intrinsic resistance of the material and the applied mechanical strain has been observed. Instead, a considerable linearity ($\mathrm{R^2}=0.96$) between the externally applied mechanical strain and the generated (piezoelectric) electric power has been found, which suggests the great potential of the latter for conducting off-the-grid strain monitoring applications.
\end{abstract}



\begin{highlights}  
\item The obtained $d_{33}$ piezoelectric factor proves 47 times greater than previous studies.
\item Homogeneous filler dispersions enhance the piezoelectricity of rGO-cement sensors.
\item Strong linearity is found between piezoelectric output power and mechanical strain.
\item A novel circuital model is proposed to simulate piezoelectric rGO-cement sensors. 
\item The piezoelectric factors are obtained by OCP-voltammetry and mechanical tests.
\end{highlights}



\maketitle

\section{Introduction}\label{Sect0}

Concrete constitutes the most widely-used construction material worldwide with a global production of $\approx$26 Gt/year~\cite{Monteiro2017}. The production of Portland cement, the main component of concrete, is estimated at 4.4 Gt/year and prospects indicate that this figure may exceed 5.5 Gt/year by 2050~\cite{lehne2018making}. The rapid development of the concrete industry followed the economic boom after World War II, and a large section of the built infrastructure was constructed during the second half of the 20$^{\textrm{th}}$ century. Nonetheless, the life-span of reinforced concrete structures is typically of 50-100 years or even less~\cite{Truong2022}, which implies that a broad section of the built stock has reached or surpassed its life expectancy. Evidence of the formidable challenge posed by the management of ageing infrastructure are some of the most recent catastrophic collapses such as the Morandi Bridge in 2018 (Genoa, Italy)~\cite{Calvi2019} of the Fern Hollow Bridge in 2022 (Pittsburg, US). This circumstance has fostered the development of new infrastructure maintenance plans and standards (see e.g.~\cite{Commission2019,GUIDA2020}), as well as considerable funding efforts devoted to R\&D actions in the realm of Structural Health Monitoring (SHM). In this context, the development of innovative micro- and nano-engineered multifunctional materials has rendered broad new possibilities in the realm of SHM. In particular, self-sensing cementitious composites (SSCCs) offer clear advantages with respect to traditional monitoring technologies (e.g.~strain gauges, optic sensors, piezoelectric ceramic, accelerometers, etc.). These innovative materials can be cast in the shape of load-bearing sensors offering high compatibility with the host structure (both through coatings or embeddable sensors, as well as distributed sensing solutions) and similar durability to standard concrete~\cite{Teomete2013, Ferreira2016, mendoza2018reinforcing}. Most research efforts have been devoted to the development of piezoresistive cement-based composites, whose self-sensing ability stems from the dependency of their electrical conductivity upon mechanical deformations. Hence, these materials allow getting information about the structural integrity and detecting damage through electrical resistivity measurements~\cite{li2019multifunctional}. The last decade has witnessed broad engineering applications of SSCCs~\cite{Han2015a}, such as strain sensing~\cite{Galao2014}, damage detection and localization~\cite{Downey2017}, structural modal identification~\cite{Ding2020}, monitoring of train loads~\cite{Ding2022}, weigh-in-motion systems~\cite{birgin2023self}, and human motion detection~\cite{DONG2022129130}, to mention a few. Nonetheless, despite the obvious motivation of these new materials, their technological transfer to routine engineering practice is anecdotal, and their applicability remains virtually exclusive to laboratory environments.



Carbon-based conductive nano-fillers are becoming increasingly popular for the development of SSCCs, including carbon fibers (CFs), carbon black (CB), carbon nanotubes (CNTs), graphene (G) and derivatives such as graphene nano-platelets (GnPs), graphene oxide (GO), or rGO~\cite{Tian2019}. When dispersed into a cementitious material, the resulting composite gains enhanced electrical conductivity, which may be of several orders of magnitude higher than that of the pristine matrix. For instance, Qureshi and Panesar~\cite{Qureshi2020} reported a reduction of 18\% in the 28-day electrical resistance of cement when adding rGO at a small concentration of 0.16 wt\%. The electrical transport properties of SSCCs are typically explained through the percolation theory accounting for two different conduction mechanisms, namely tunnelling and conductive networking~\cite{Chang2009}. The percolation threshold indicates the critical filler volume fraction at which the doping particles touch one another, originating continuous conductive paths with the subsequent drastic increase of the overall conductivity. Below percolation, the transfer of electrons takes place through the (dielectric) matrix and the potential barrier between non-contacted fillers in virtue of a quantum tunnelling mechanism. Above percolation, both mechanisms act simultaneously, although conductive networking becomes dominant. In this light, the self-sensing property of SSCCs can be understood as the alteration of these mechanisms driven by mechanical strains.

The origin of the strain self-sensitivity of SSCCs has been ascribed in the literature to three different mechanisms~\cite{Ramachandran2022}: (i) piezoresistivity, (ii) piezopermittivity, and (iii) piezoelectricity. \textit{Piezoresistivity} constitutes the most extensively investigated mechanism. It is defined as the one-way coupling between the material electrical resistivity and mechanical strains~\cite{Dong2019}, that is to say, mechanical loads affect the electric field but not the opposite. Jang \textit{et al}.~\cite{Jang2022} investigated the electrical and piezoresistive properties of cement-based sensors doped with multi-walled CNTs (MWCNTs) and CB, respectively, exposed to various temperatures. Their results reported gauge factors (GFs) in the range $\approx$18-50 for temperature values between 25 and 400$^\circ$C and MWCNTs contents of 0.2, 0.5, and 1 wt\%. Similarly, Dong and co-authors~\cite{DONG2019107488} investigated the piezoresistive properties of cement doped with 3 wt\% CB under various temperature and water content conditions. Their findings revealed that the piezoresistive properties were not significantly affected by thermal variations in the range of -20 to 100$^\circ$C, once the heat gradients between the sensors and the working environment were removed. Conversely, the water content had a more significant impact, with optimal sensitivities achieved at a content of around 8\%, resulting in a GF of 488. Xu and Zhang ~\cite{xu2017pressure} and Guo \textit{et al}.~\cite{guo2021study} reported GFs in the ranges 0.08-7.78 (0.67-2.00 wt\%) and 3-16 (0.05-0.15 wt\%) for cement doped with GnP and GO, respectively. Both experimental and theoretical evidence in the literature have shown that increasing compressive strains lead to decreases in the electrical resistivity of the composite and vice-versa for tensile strains. Other recent studies have focused on analyzing the piezoresistive properties of rGO/cement composites under conditions of sulphuric acid attack~\cite{DONG2021100563} and dynamic mechanical excitation~\cite{QI2023130340}. \textit{Piezopermittivity} describes the effect of mechanical strains upon the dielectric behaviour of a material. In general, the capacitance of cementitious materials stems from the movement of charge carriers under an electric field. Specifically, electric fields induce the ions in the medium to be repelled causing a dipolar effect, thus resulting in polarization. When subjected to a constant potential difference, polarization in cement-based materials manifests as electric current decreases over time as a result of capacitive effects~\cite{wen2001electric,Cao2004}. With respect to the relevance of piezocapacitive effects in the electromechanical response of SSCCs, disparate conclusions have been reported in the literature. For instance, Han \textit{et al}.~\cite{Han2012} found no significant variation in the internal capacitance of cement composites doped with low filler contents of MWCNTs (0.1 and 0.5 wt\%) under low compressive loadings ($\leq$6 MPa). Conversely, Hou and co-authors~\cite{Hou2022} reported comparable GFs in terms of internal resistance and capacitance. Those authors investigated the electrical properties of cement doped with steel fibers at concentrations between 0.5 and 2 vol.\% through DC and AC resistivity measurements and large compressive loads (0-25 MPa). Their results evidenced that capacitance-based measurements provide noisier signals and slightly lower strain sensitivities compared to resistance-based measurements. Finally, \textit{piezoelectricity} relates the property of certain materials to produce an external voltage when a mechanical load is exerted on it~\cite{Elahi2018}. Unlike piezoresistivity, piezoelectricity represents a two-way coupling where mechanical strains affect the electric field and vice-versa. The intrinsic piezoelectric properties of cement hydrates without functional fillers have been long recognised in the literature~\cite{Sun2004,Shi2018}. It is believed that the appearance of permanent dipoles in cement-based materials originating piezoelectricity stems from the interaction of mobile charge carriers (e.g. the ions in the pore solution) with the atoms or ions in the solid phase, particularly those associated with certain micro-structural elements such as the diffuse interfaces in cement paste. However, typical piezoelectric coefficients of hardened cement are in the order of 1.0E-8 A/MPa, which is too small to be exploited for monitoring purposes. For this reason, researchers have strived to increase the piezoelectricity of cementitious materials, primarily through the addition of piezoelectric functional fillers such as lead zirconium titanate (PZT) or polyvinylidene fluoride (PVDF)~(see e.g.~\cite{Jaitanong2014,Xu2022}). In this light, the incorporation of conductive carbon-based nanofillers has been mainly investigated for the purpose of reducing the polarization voltage of the resulting three-phase composite and enhancing its effectiveness. For instance, Gong and co-authors~\cite{Gong2010} reported that the addition of 1.0 vol.\% CB to PZT/cement allowed achieving an efficient polarization at 40 kV/cm and room temperature, attaining an effective piezoelectric strain constant $d_{33}$ of 28.5 pC/N. Similar experiences in the literature concerning the use of CNTs, CB or GnPs (refer to e.g.~\cite{Huang2009,Potong2017,Jaitanong2018}) report the achievement of considerably high piezoelectric strain constants $d_{33}$ within the range 13-122 pC/N. Overall, SSCCs exploiting piezoelectricity require no power supply (neither DC nor AC) to conduct electrical measurements, which represents a clear practical advantage with respect to piezoresistive SSCCs. This feature may effectively boost the technological transfer of SSCCs towards their wide-scale implementation, since the provision of power supply to civil engineering structures at a regional level is simply infeasible. Nonetheless, the low toughness and the health and environmental concerns related to piezoelectric ceramics represent a major limitation for their practical use to develop load-bearing SSCC sensors. Alternatively, recent research works have proposed the use of DC electric fields to polarize the ions inside the fresh cement paste while curing, so enhancing its intrinsic piezoelectric properties without piezoceramics (see e.g.~\cite{Dong2007,Yaphary2016}). In this line, a noteworthy contribution was made by Al-Qaralleh~\cite{AlQaralleh2022} who reported considerable enhancements in the piezoelectric properties of cements polarized while curing at a DC potential of 5 V. Specifically, that work reported a piezoelectric voltage constant $g_{33}$ of 9.7E-5 mV m/N, a value considerably superior to that of untreated cement (3.7E-5 mV m/N). Nonetheless, whilst promising, the use of polarization techniques in the fabrication of SSCCs may limit its scalability. From the literature review above, it is clear that the use of carbon-based functional fillers to directly enhance the intrinsic piezoelectric properties of cementitious materials without the addition of piezoceramic inclusions remains scarcely explored.

Experimental research has been accompanied by considerable efforts devoted to the development of theoretical models capable of interpreting the underlying physical phenomena governing SSCCs. Overall, theoretical models reported in the literature comprise mean-field homogenization (MFH) techniques, numerical simulation, and lumped circuital models. Mean-field homogenization offers a powerful analytical or semi-analytical framework to estimate the effective physical properties of SSCCs by exploiting the statistical averaging of the contribution of the constituent phases within a representative volume element (RVE)~\cite{GarciaMacias2020}. In this regard, it is worth noting the work by Takeda \textit{et al}.~\cite{Takeda2011} who presented a micromechanics model of the electrical conductivity of CNTs/polymer composites, which allowed to distinguish the contribution of tunnelling-type contacts and conductive networking, as well as to simulate percolation and filler waviness. Another prominent contribution was made by Seidel and Lagoudas~\cite{Seidel2009} who proposed an Eshelby-Mori-Tanaka model for the study of the individual influence of the conductive mechanisms on the overall electrical conductivity of CNTs/polymer composites. Tallman and Wang~\cite{Tallman2013} extended the formulation in \cite{Takeda2011} to simulate the piezoresitive properties of CNTs-based composites under arbitrary strain states. The Eshelby-Mori-Tanaka model was also adopted by Feng and Jiang~\cite{Feng2015} and Garc\'{i}a-Mac\'{i}as \textit{et al}.~\cite{microCNTcement,GarciaMacias2018} to simulate the piezoresistivity of CNTs-reinforced composites. In those works, the piezoresistivity properties were simulated through three different mechanisms driven by mechanical strains: (i) volume expansion, (ii) filler reorientation and (iii) changes in the percolation threshold. Closed-form solutions to the formulation in reference~\cite{GarciaMacias2018} were later derived by Buroni and Garc\'{i}a-Mac\'{i}as~\cite{Buroni2021} through generalized spherical harmonics series expansions, allowing to obtain analytical expressions of the effective conductivity and piezoresistive coefficients of self-sensing composites doped with rod-like inclusions. Recently, Triana-Camacho and co-authors~\cite{TrianaCamacho2022} reported the development of an effective medium model to compute the impedance properties of CNTs/cement composites. While MFH offers a formal framework to relate the physical mechanisms governing the electrical transport properties of SSCCs, such modelling approaches may become overly simplistic for certain applications and have limited applicability to simulate electric time-variant systems. Numerical simulation methods instead allow to faithfully represent the microstructure of SSCCs and the interaction between their constituent phases. Numerical approaches primarily include molecular dynamics simulation (MDS), atomistic-continuum modelling, and numerical homogenisation models (refer to references~\cite{Qiu2021,Zhao2021} for a comprehensive state-of-the-art review). Despite such approaches offer high simulation accuracy, their considerable computational burden hinders their applicability to meso- and macro-scale systems. Alternatively, equivalent lumped-circuit models are being imposed as they provide a feasible and tractable framework to simulate electric time-varying systems. A noticeable example is the work by Kang \textit{et al}.~\cite{Kang2006a} who performed electrochemical impedance spectroscopy (EIS) testing to characterize the electrical properties of SWNT/PMMA sensors. On this basis, those authors proposed a modified Randle's circuit to represent the dynamic behaviour of the sensors. Similarly to other works, the strain sensing response was enabled by a linear relationship between the variation of the relative change of the internal electrical resistance and the axial strain. This type of approaches neglects any kind of non-linearity in the response, thereby its applicability typically limits to small deformation ranges and filler contents sufficiently far from the percolation threshold. A similar study is the one by Loh \textit{et al}.~\cite{Loh2008} on SWNT-PSS/PVA thin films, who formulated an RC-circuit model by inverse calibration from electrical impedance spectroscopy measurements. Good agreements with experimental data were obtained by defining the contact resistance, the double layer capacitance and the bulk resistance as exponentially decaying functions with time. Sanli \textit{et al}.~\cite{Sanli2016} proposed an RC equivalent circuit based on the impedance response of CNTs/epoxy films. An interesting aspect of that work regards the consideration of not only the dependence of the internal resistance of the circuit on external strain, but also strain-dependent capacitive effects originating piezocapacitance. Another noteworthy contribution was made by Materazzi \textit{et al}.~\cite{Materazzi} who reported an experimental campaign devoted to the assessment of the changes in the electrical resistance of SSCCs doped with CNTs under the action of sinusoidal compression loads. Their analyses showed that the amplitude of the electrical resistance variation increases with the frequency of excitation. Following that work, D'Alessandro \textit{et al}.~\cite{DaAlessandro2014} proposed a Randle's equivalent circuit-based electromechanical model consisting of two resistors and a capacitor, accounting for the contact resistance (cables and electrodes), electric polarization and internal dissipation effects. An important conclusion of that work was that the dynamic response of CNTs-based SSCCs is not monochromatic, but rather contains superharmonics. However, while the presence of superharmonics was anticipated by the model, the rising amplitude of the response with increasing frequency could not be reproduced. Garc\'{i}a-Mac\'{i}as \textit{et al}.~\cite{garcia2017enhanced} extended the circuital model proposed in \cite{DaAlessandro2014} considering piezoresitive, piezocapative, and piezoelectric effects. The reported results demonstrated that the amplification of the electric resistance of CNTs-based SSCCs with the frequency of excitation under DC power supply can only be explained by a piezoelectric effect, which was further confirmed by specific experiments. Another noticeable result of that work regards the low capacitance GFs, which agrees with the experimental evidence previously reported by Han \textit{et al}.~\cite{Han2012}.

As revealed by the literature review above, although significant efforts have been exerted on the characterization of piezoresistive SSCCs (e.g. refer to~\cite{DONG2021100563,QI2023130340}), comparatively little attention has been devoted to developing and harnessing piezoelectric rGO/cement composites for strain sensing applications. Nonetheless, SSCCs relaying on the piezoresistive principle require an external power supply to conduct strain measurements, which poses a major practical limitation for civil engineering structures, often located in remote areas with no access to an electrical grid. In this light, this work presents the manufacture and characterization of highly piezoelectric rGO/SSCCs for the development of self-powered strain sensors for SHM applications. The self-powered ability of such sensors lays on the fact that they generate their own sense signal without requiring an external power supply. To this aim, a comprehensive experimental methodology is proposed to characterize the piezoelectric coupling coefficients $g_{33}$ and $d_{33}$. The methodology comprises cyclic voltammetry, open circuit potential determination (OCP), and uniaxial quasi-static compression tests. The developed methodology has been applied to laboratory samples manufactured following two different filler dispersion methods. Furthermore, a novel lumped circuital model involving piezoresistive, piezocapacitive, and piezoelectric sensing principles is developed and experimentally validated. The presented results demonstrate that samples prepared by ultrasonic cleaner dispersion achieve optimal properties, with piezoelectric coefficients about 47 times greater compared to previously reported composites in the literature. Overall, the presented results evidence the potential of rGO/cement composites to develop self-powered sensors for SHM applications, for which the proposed circuital model may constitute a powerful signal processing tool allowing to translate the electrical output into strain signals. To the best of the authors' knowledge, this work presents the first evidence in the literature of the use of piezoelectric rGO/SSCCs for mechanical strain sensing applications requiring no external power supply.

The paper is organized as follows. Section~\ref{Sect01} describes the manufacturing process and the testing details. Section~\ref{Sect1} reviews the formulation of previously reported equivalent lumped circuits and presents the newly proposed model. Section~\ref{Sect2} presents the numerical results and discussion and, finally, Section~\ref{Sect3} concludes the paper.


\section{Methods and Materials}\label{Sect01}

This section presents the materials, manufacturing process and testing methods adopted in this work. The manufacturing process of the specimens followed two different dispersion methodologies, and the resulting dispersions were later characterized by Dynamical Light Scattering (DLS). Afterwards, the effective piezoelectric properties were characterized by cyclic electromechanical tests. Finally, this section also presents the theoretical formulation of the developed equivalent lumped circuit to replicate the electromechanical response of piezoresistive/piezocapaticitive/piezoelectric rGO/cement composites.


\subsection{Materials and preparation of specimens}\label{fabrication}

Bulk rGO powder (5-10 layer flakes with average thickness around 500 $\upmu$m) was obtained in the ``\textit{Laboratorio de Espectroscopia Atómica y Molecular}'' (LEAM) of the ``\textit{Universidad Industrial de Santander}'' (Colombia) following the modified Hummers’ method set out by Arenas and Marcano \textit{et al.}~\cite{arenas2019methodology, Marcano2010}. Other materials used in the admixtures included standard Portland cement produced under the Colombian standard NTC121 (equivalent to ASTM C1157); ultrapure water from a Milli-Q IQ 7000 equipment; and 2.0 mm copper desoldering wires used as embedded electrodes. 

\begin{figure}[ht]
    \centering
    \includegraphics[scale=0.8]{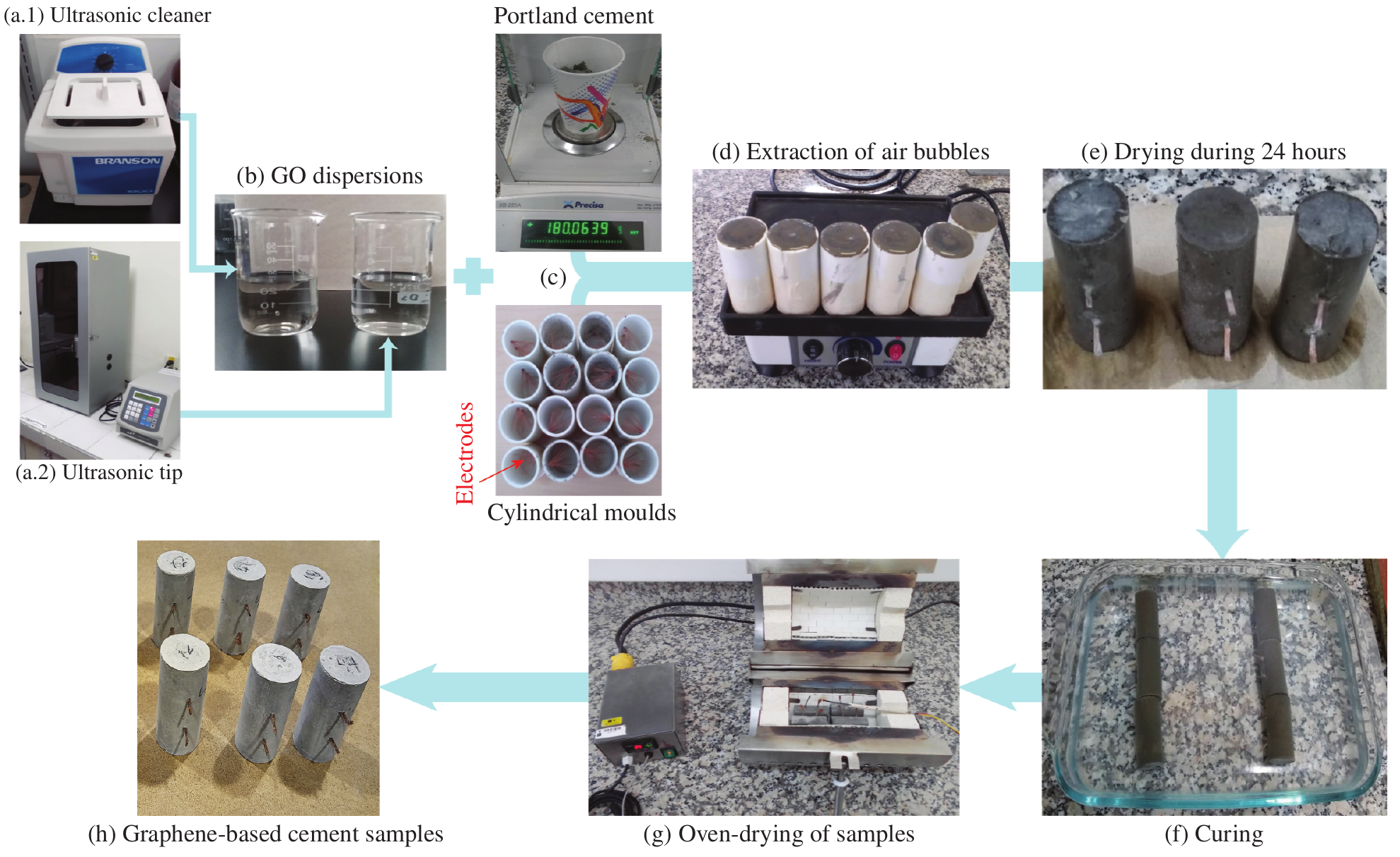}
    \caption{Fabrication stages of rGO-based cement composites.}
    \label{fabrication_samples}
\end{figure}

The fabrication process of the specimens is schematically illustrated in Fig.~\ref{fabrication_samples}. In the first place, 0.24 mg of rGO powder is added to 80 mL of ultrapure water. Separately, two mixtures are dispersed for 2 h using an ultrasound cleaner machine model Branson 1800 (a.1) with a working frequency of 40 kHz. Simultaneously, the other two mixtures are dispersed using an ultrasonic tip set (a.2) with an energy of 490 J and amplitude of 40 \% applied during 30 min with (on/off) cycles of (20 s/20 s), respectively, following the same prescriptions indicated by Echeverry-Cardona \textit{et al.}~\cite{echeverry2021} for CNTs-based composites. The two different dispersion techniques used in this work, namely (i) ultrasonic cleaner and (ii) ultrasonic tip set, are hereafter referred to as Method 1 and Method 2, respectively. For each dispersion method, 180 mL of rGO/water solutions are prepared (b). Afterwards, Portland cement is mixed with the rGO/water solutions with a water-to-cement ratio (w/c) of 0.47, achieving rGO dispersions of 0.14\% by cement mass. Note that such a filler content is far below the percolation threshold, which has been reported in the literature to be around 2.0 wt.\% (see e.g.~\cite{al2016electrical,zhang2020experimental}). Then, the admixture is poured into cylindrical molds (c) which contain copper desoldering wires arranged forming two parallel plane electrodes. The electrodes are fixed into the molds using two holes with a separation distance of 20 mm. Paper tape is used to secure the attachment between the outgoing electrodes and the mold surface. In particular, 3 cylindrical specimens (labelled with s$i$-M$j$ with $i=1,\ldots,3$ and $j=1,2$ denoting the $i$-th sample number and the $j$-th dispersion method) with diameters of 30 mm and lengths of 60 mm are prepared for each dispersion method according to the ASTM C349-18 standard~\cite{astm349}. In this light, the repeatability of the material is statistically assessed through the analysis of variance (ANOVA) test (significance level of 0.05) of the three samples. Then the moulds are put on a vibration table for 10 min to minimize the presence of air bubbles in the samples (d). Once finished, the specimens are left to dry at room temperature for 24 h (e), followed by a curing period of 28 days in ultrapure water (specific resistance of 18 M$\mathrm{\Omega}$) (f). When the curing stage is completed, the samples are further dried in an oven at 40 $^\circ$C for 24 h as similarly adopted in~\cite{echeverry2021} (g), after which the specimens are ready for electromechanical characterization (h). 

\begin{figure}[ht]
\centering
\includegraphics[scale=1.0]{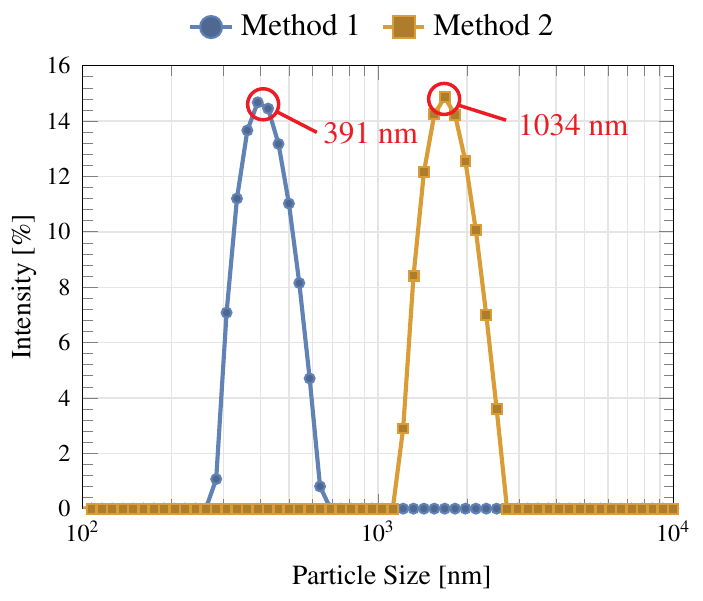}
\caption{Particle size distributions of rGO dispersed in 1:100 water for dispersion Methods 1 and 2.}
\label{ps}
\end{figure}

It is important to remark that the particle size distribution of rGO is directly determined by the dispersion methodology and the sonication time as previously reported by other authors such as Kiamahalleh \textit{et al.}~\cite{VALIZADEHKIAMAHALLEH2020118832}. The quality of the dispersion determines the electromechanical behaviour of the resulting composite, being in general desirable to avoid the formation of filler clusters acting as microstructural defects. The rGO dispersions obtained by the two different methodologies described above were characterized by DLS at a 90$^\circ$ angle using a Litesizer 500 particle analyzer from Anton Paar. To do so, the dispersions were diluted in water in a proportion of 1:100 and later poured into quartz cuvettes. The particle analyzer internally tests the dispersions in triplicate, each one obtained as the average between 6 measurements. The weighted intensity versus particle size curves obtained by dispersion Methods 1 and 2 are show in Fig.~\ref{ps}. It is noted in this figure that Method 1 yielded the smallest particle size with a mean value of 391 nm (intensity peak of 14.68\%), considerably smaller than the one obtained by Method 2 with a mean particle size mean of 1681 nm (intensity peak of 14.86\%). This is conceivably ascribed to the fact that, although ultrasonic tip dispersion (Method 2) generates denser energy compared to the ultrasonic cleaner (Method 1), high sonication energies may increase the charge density by the oxygen functional groups and induce the reaggregation of rGO sheets as shown in~\cite{VALIZADEHKIAMAHALLEH2020118832} and, thus, result in larger particle sizes. It is important to remark that close agreements can be found between the average particle size obtained by Method 1 in this work and previously reported results by Wang and co-authors~\cite{wang2017effect}. 

\subsection{Experimental assessment of piezocapacitive/piezoelectric properties}\label{piezo_char}

\begin{figure}[ht]
\centering
\includegraphics[scale=0.9]{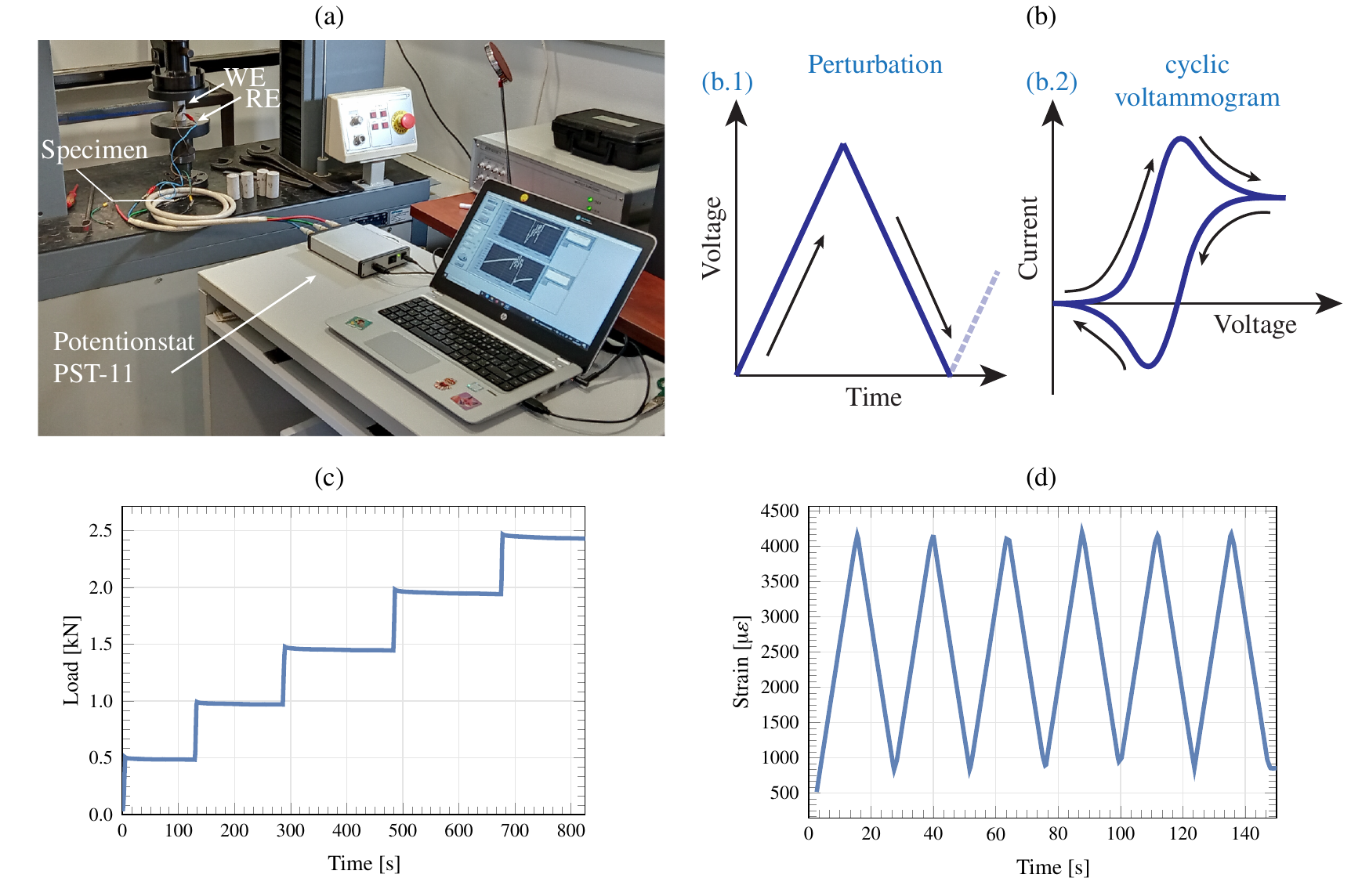}
\caption{Experimental setup to characterize the electromechanical properties of rGO-based cement composites (a), and schematic representation of cyclic voltammetry (b). Loading conditions: stair-step (c) and triangular cyclic loading (d).}
\label{experimental_setup}
\end{figure}

The mechanical response of the manufactured specimens was characterized through displacement-controlled uni-axial compression tests with a universal testing machine Servosis ME 405. In parallel, the electrical behaviour of the samples was characterized using a Potentiostat PST-11 from DinTech manufacturer according to the experimental set-up shown in Fig.~\ref{experimental_setup} (a). The two electrodes of the specimens were connected to the reference electrode (RE) and the working electrode (WE), while the ground was connected to the grip of the universal testing machine. The data extracted from the universal testing machine and potentiostat PST-11 were synchronized in time and processed using Python scripts written in Google Colab. 

A novel methodology involving cyclic voltammetry and two sets of loading conditions was developed to assess the piezocapacitive and piezoelectric properties of rGO/cement composites. Firstly, the specimens were subjected to stair-step loading with compression loads increasing from 50 kg to 250 kg, as shown in Fig.~\ref{experimental_setup} (c). For each step in the loading sequence, cyclic voltammetry was performed to estimate the capacitive properties of the samples and their dependence on the applied strain. Cyclic voltammetry is a popular electroanalytical technique to study the electrochemical properties of analytes. This technique consists in applying a cyclic voltage with a triangular waveform (Fig.~\ref{experimental_setup} (b.1)) to the WE with respect to the RE and measuring the produced electric current (potentiostatic). The results of cyclic voltammetry are typically presented in the shape of current–potential curves (cyclic voltammograms) as sketched in Fig.~\ref{experimental_setup} (b.2). In this light, the capacitance $C$ of the samples can be obtained as:

\begin{equation}\label{cap}
    C = \frac{1}{2 \, s_r \, \Delta v }\int i(v) \, \textrm{d}v,
\end{equation}

\noindent where $i$ is the electrical current, $v$ the voltage, $s_r$ the scan rate, and $\Delta v$ the voltage window, i.e., the amplitude of the applied voltage waveform. In this work, $v$ spans from -0.5 V to 0.5 V, that is $\Delta v = 1$ V. It is important to remark that the increases in the electrical current and the voltage difference increment also depend upon the scan rate, manifesting the highest pseudocapacitance in a rGO-based composite when cyclic voltammetry is set below 50 mV/s as previously reported by Czepa and co-authors~\cite{czepa2020reduced}. For this reason, the electrical characterization was conducted in this work with a scan rate of $s_r=25$ mV/s. 

Afterwards, the samples were subjected to triangular cyclic loads with a fixed displacement rate of 1 mm/min (i.e.~0.05 Hz) (see Fig.~\ref{experimental_setup} (d)) to measure their piezoelectric response. In these tests, the correlation between the generated voltage and mechanical forces was estimated through linear regression. The estimated correlations (units of V/N) were multiplied by the conversion factor $A/s_e$ to estimate the voltage piezoelectric coefficient $g_{33}$, with $A$ and $s_e$ being the cross-section area of the specimens and the distance between the electrodes, respectively. Note that this conversion factor allows transforming voltage into electric field and the mechanical load into stress as suggested by Al-Qaralleh~\cite{AlQaralleh2022}. Then, the correlation between the generated electric field and the corresponding stress represents the voltage piezoelectric coefficient $g_{33}$, which can be related to the piezoelectric charge coefficient $d_{33}$ as follows:

\begin{equation}\label{eq:d33}
    d_{33} = C \, \frac{s_e}{A} \, g_{33}.
\end{equation}

\section{Equivalent piezoresistive/piezocapacitive/piezoelectric lumped circuit}\label{Sect1}

\begin{figure}[ht]
\centering
\includegraphics[scale=0.85]{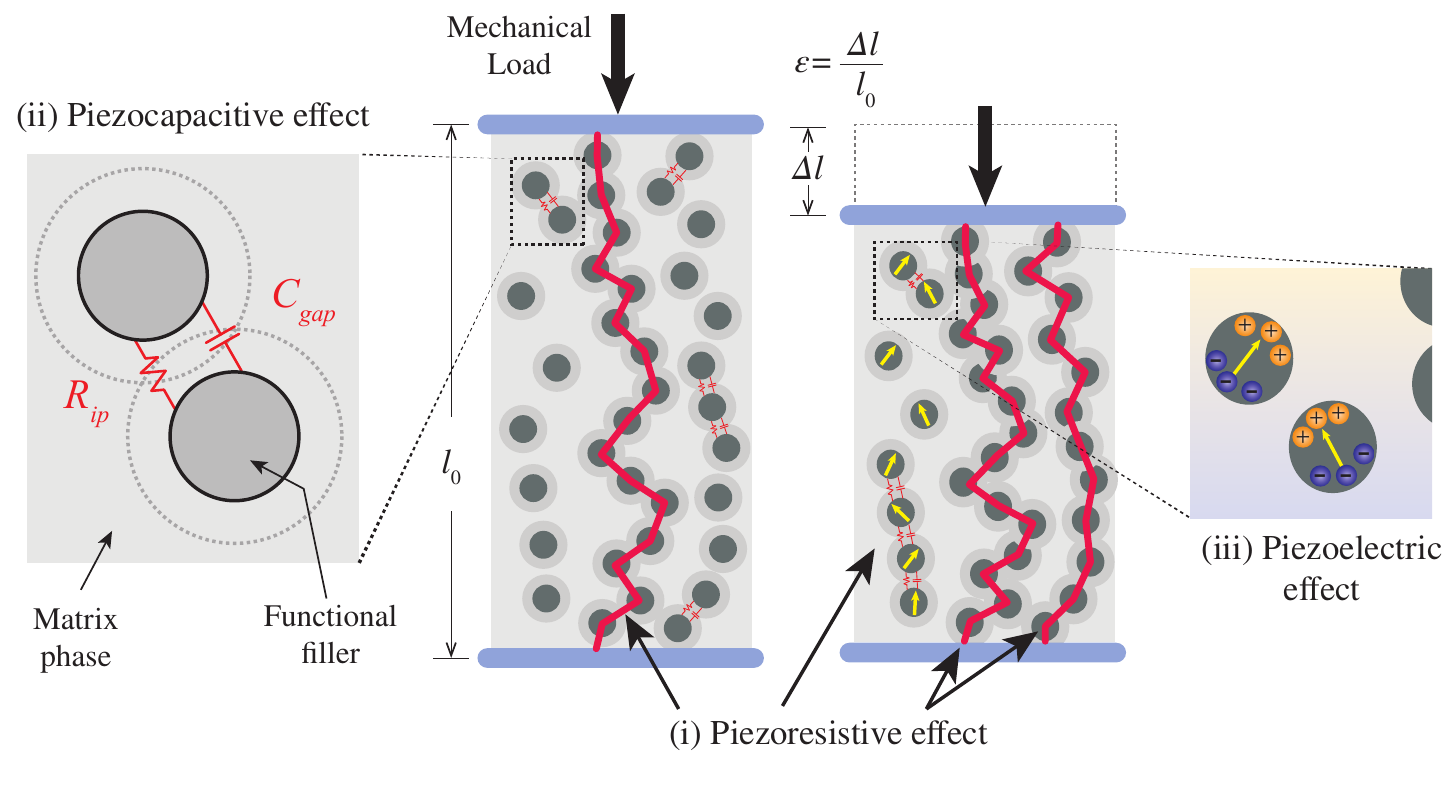}
\caption{Schematic representation of the physical phenomena governing the electric transport properties of rGO/cement composites: (i) piezoresistivity, (ii) piezocapacitance, and (iii) piezoelectricity.}
\label{wp}
\end{figure}

The main electronic transport mechanisms hypothesized in this work to describe the electromechanical behaviour of rGO-based cement composites are sketched in Fig.~\ref{wp}. These account for three different strain-sensing principles, namely (i) piezoresistivity, (ii) piezocapacitance, and (iii) piezoelectricity. As anticipated above, the electrical conductivity of carbon-based composites has been extensively ascribed to the coupled contribution of electron hopping and conductive networking mechanisms. In this light, (i) piezoresistivity arises as a result of three main strain-driven mechanisms: volume expansion, filler reorientation and, changes in the tunnelling resistance (see e.g.~\cite{GARMA2016STR}). Firstly, the effects of filler reorientation and changes in the tunnelling resistance have been shown to be dominant for filler contents below the percolation threshold~\cite{GarciaMacias2018}, as it is the case in the present work. Tunnelling effects arise among particles with a separation less than a certain cut-off distance ($\approx$0.5 nm for cementitious media~\cite{xu2010modeling}). Therefore, as the material experiences compression, the inter-particle distance decreases, increasing the amount of charges flowing through quantum tunnelling effects and so the overall electrical conductivity. Instead, strain-induced volume expansion effects become dominant for filler contents far from the percolation threshold. Since carbon-based fillers are often orders of magnitude stiffer than the matrix, the deformation is primarily sustained by the matrix phase, which leads to variations in the apparent filler volume fraction~\cite{Feng2014}. As the volume decreases under compression, the inter-particle distance decreases rising the likelihood of forming new conductive paths and increasing the effective electrical conductivity. These phenomena explain the frequently observed piezoresistive behaviour in SSCCs in which compressive strains lead to increases in the overall electrical conductivity, and vice versa for tensile strains (see e.g.~\cite{madbouly2020evaluating}). The origin of (ii) piezocapacitive effects can be intuitively explained by the insert in Fig.~\ref{wp}. In general, there exists a dielectric gap between every two conductive particles which can be conceived as a RC system with $R_{gap}$ and $C_{gap}$ corresponding to the contact resistance and capacitance, respectively. Under the action of a compression load, the distance between the particles is on average reduced, which will result in a decrease of $R_{gap}$ and an increase in the $C_{gap}$ value. Hence, in general, compression loads will lead to decreases in the electrical resistance and decreases in the overall capacitance. Finally, as sketched in Fig.~\ref{wp}, the origin of (iii) piezoelectric effects in rGO/cement composites is ascribed to the polarization of the sets formed by rGO sheets and cement hydration products such as calcium silicates~\cite{LV2013}. 

\begin{figure}[ht]
    \centering
    \includegraphics[scale=1.0]{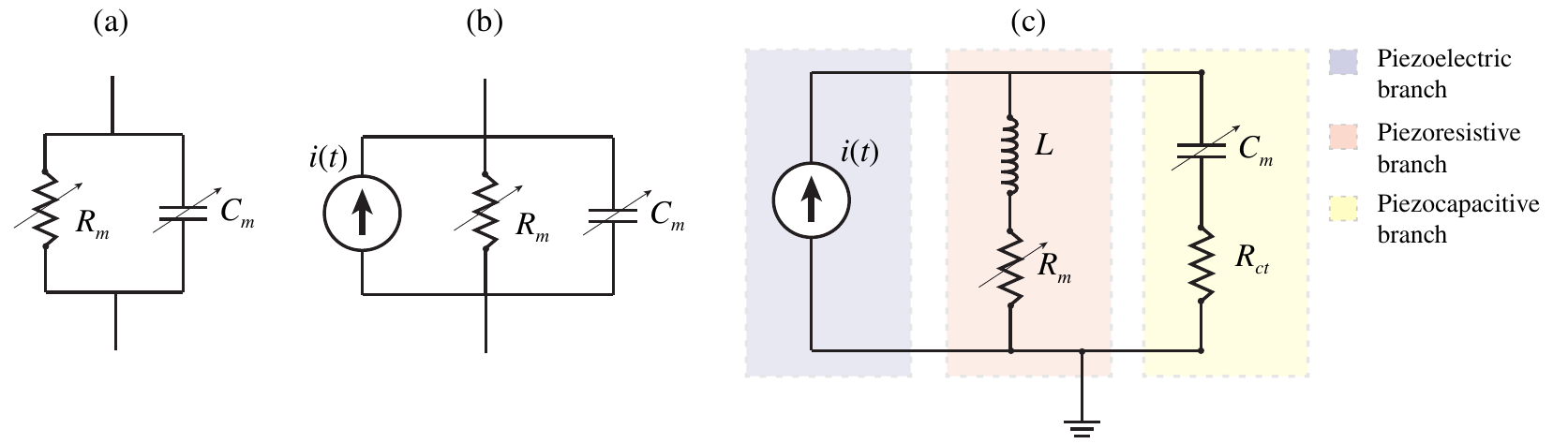}
    \caption{Piezoresistive/piezocapacitive unit sensing cell (a). Piezoresistive/Piezocapactitive/Piezoelectric sensing cell (b). Proposed equivalent lumped circuit model to simulate the electrical response of piezoresistive/piezocapacitive/piezoelectric rGO-based cement composites (c).}
    \label{lcm}
\end{figure}

The most commonly adopted circuital model to describe the electrical behaviour of SSCCs is the modified Randle's circuit depicted in Fig.~\ref{lcm} (a). This model represents the active volume of the specimen (portion of the material in between the electrodes) as a capacitor $C_{m}$ and a resistor $R_{m}$ in parallel. In this light, piezoresistivity and piezocapacitance can be simulated through the strain-dependent definition of $R_{m}$ and $C_{m}$, respectively. Generally, the relationship between the relative change in the internal electrical resistance and the axial mechanical strain $\varepsilon(t)$ (negative in compression) is assumed to be linear:

\begin{equation}
\lambda_{R_m} = \frac{\frac{\Delta R_{m}}{R_{m}^o}}{\varepsilon(t)},
\label{eq:FCR}
\end{equation}

\noindent where $\lambda_{R_m}$ denotes the gauge factor of the sensor, $R_{m}^o$ is the unstrained internal resistance, and $t$ denotes the time variable. Similarly, piezocapacitive effects can be simulated through a second gauge factor, $\lambda_{C_m}$, as~\cite{Sanli2016}:

\begin{equation}
\lambda_{C_m} = -\frac{\frac{\Delta C_{m}}{C_{m}^o}}{\varepsilon(t)},
\label{eq:FCC}
\end{equation}

\noindent with $C_{m}^o$ denoting the unstrained capacitance. 

The contribution of piezoelectric effects can be incorporated in the shape of a strain-dependent current source as sketched in Fig.~\ref{lcm} (b) as previously proposed by Garc\'{i}a-Mac\'{i}as \textsl{et al.}~\cite{garcia2017enhanced}. Assuming the linear piezoelectricity theory holds, the electric current $i(t)$ introduced in the circuit is directly related to the time derivative of the applied strain as:

\begin{equation}\label{piezoelec1}
i(t)=S_q\frac{\textrm{d}\varepsilon(t)}{\textrm{d}t},
\end{equation}

\noindent with $S_q$ being a linear piezoelectricity coefficient. A similar relationship has been also suggested by several authors for other cement-based composites (see e.g.~\cite{SHUAI2020108564}). Nevertheless, the electric current measurements obtained from the rGO/cement composites manufactured in this work did not exhibit a clear linear relationship with the applied strain. These results, as shown later in Section~\ref{Sect22}, may indicate that linear piezoelectricity theory may not apply to this material.

From the previous discussion, a novel lumped circuit is proposed in this work to simulate the electromechanical response of rGO-based cement composites. The model, shown in Fig.~\ref{lcm} (c), was inspired by the Butterworth-Van Dike circuit, which is commonly used to describe the piezoelectric effects in ceramics (see e.g.~\cite{Kim2008}). The proposed model accounts for three different branches in parallel representing the piezoelectric, piezoresistive, and piezocapacitive contributions. The piezoelectric branch is defined by a general current source $i(t)$ with no assumptions upon its dependency with the applied mechanical strain. Discussion on the origin and definition of this term are reported hereafter in Section~\ref{Sect2}. The piezoresistive branch is represented by a linear variable resistance $R_m(\varepsilon)$ in series with an inductor $L$. The latter allows reproducing the electrical damping effects arising as a result of the appearance of current loops among the rGO sheets. Finally, the piezocapacitive branch incorporates a linear variable capacitor $C_m(\varepsilon)$ and a resistor $R_{ct}$ representing the variable internal capacitance and the charge transfer resistance, respectively.  

Let us denote with $V_i(t)$ the voltage drops across the $i$-th element in the circuit model in Fig.~\ref{lcm} (c), with sub-index $i$ indicating the particular element. In this light, making use of Kirchhoff's voltage law, that is $V_{R_m}(t) + V_{L}(t) = V_{R_{ct}}(t) + V_{C_m}(t)$, it is possible to extract the governing differential equation of the proposed lumped circuit as:

\begin{equation}
R_m(\varepsilon) \frac{\textrm{d} i_L(t)}{\textrm{d}t} + \frac{\textrm{d}R_m(\varepsilon)}{\textrm{d}t} i_L(t) + L\frac{\textrm{d}^2 i_L(t)}{\textrm{d}t^2} = R_{ct}\frac{\textrm{d} i_C(t)}{\textrm{d}t} + \frac{i_C(t)}{C_m(\varepsilon)},
    \label{voltagesum}
\end{equation}

\noindent with $i_L(t)$ and $i_C(t)$ denoting the electric currents crossing the inductor and the piezocapacitive branch, respectively. Then, combining Eq.~(\ref{voltagesum}) with the Kirchhoff's current law $i(t) = i_C(t) + i_L(t)$, the differential equation in terms of current $i_C(t)$ can be obtained as follows:

\begin{equation}
L\frac{\textrm{d}^2 i_C(t)}{\textrm{d}t^2} + \left(R_m(\varepsilon) + R_{ct}\right) \frac{\textrm{d} i_C(t)}{\textrm{d}t} + \left(\frac{\textrm{d}R_m(\varepsilon)}{\textrm{d}t}+\frac{1}{C_m(\varepsilon)}\right) i_C(t) = L\frac{\textrm{d}^2 i(t)}{\textrm{d}t^2}+R_m\frac{\textrm{d} i(t)}{\textrm{d}t}+\frac{\textrm{d}R_m(\varepsilon)}{\textrm{d}t} i(t).
    \label{edo}
\end{equation}

Introducing the strain dependency of $R_m$ and $C_m$ from Eqs.~(\ref{eq:FCR}) and (\ref{eq:FCC}), Eq.~(\ref{edo}) can be rewritten after some manipulation as: 

\begin{equation}
L\frac{\textrm{d}^2 i_C(t)}{\textrm{d}t^2} + \left[R_m^o\left(1+\lambda_{R_m}\varepsilon(t)\right) + R_{ct}\right] \frac{\textrm{d} i_C(t)}{\textrm{d}t} + \left[R_m^o\lambda_{R_m}\frac{\textrm{d} \varepsilon(t)}{\textrm{d}t}+\left[C_m^o\left(1+\lambda_{C_m}\varepsilon(t)\right)\right]^{-1} \right] i_C(t) = I(t),
    \label{edofinal}
\end{equation}

\noindent with

\begin{equation}
I(t) = L\frac{\textrm{d}^2 i(t)}{\textrm{d}t^2}+R_m^o\left(1+\lambda_{R_m}\varepsilon(t)\right)\frac{\textrm{d} i(t)}{\textrm{d}t}+R_m^o\lambda_{R_m}\frac{\textrm{d} \varepsilon(t)}{\textrm{d}t} i(t).
\end{equation}

Note that Eq.~(\ref{edofinal}) is a second-order ordinary differential equation with variable coefficients that needs to be solved numerically. In this work, the proposed circuit has been implemented in Matlab Simulink environment, and the Runge-Kutta integration method has been used to extract numerical solutions to Eq.~(\ref{edofinal}). Once the capacitive current $i_C(t)$ is obtained, the voltage generated by the sensor can be computed as:

\begin{equation}
V(t) = C_m^o\left(1+\lambda_{C_m}\varepsilon(t)\right)\frac{\textrm{d}i_C(t)}{\textrm{d}t} + R_{ct}i_C(i).
\label{ouputvoltage}
\end{equation}

\section{Results and discussion}\label{Sect2}

The estimation of the piezoelectric coefficients $g_{33}$ and $d_{33}$ through cyclic voltammetry is firstly presented in Section~\ref{Sect21}. Afterwards, Section~\ref{Sect22} reports the experimental characterization of the electromechanical response of the manufactured rGO/cement composites under cyclic quasi-static compressive loading. Finally, Section~\ref{Sect23} reports the calibration of the lumped equivalent circuital model previously introduced in Section~\ref{Sect1}.

\subsection{Piezoelectric coefficients}\label{Sect21}

\begin{figure}[ht]
\centering
\includegraphics[scale=1.0]{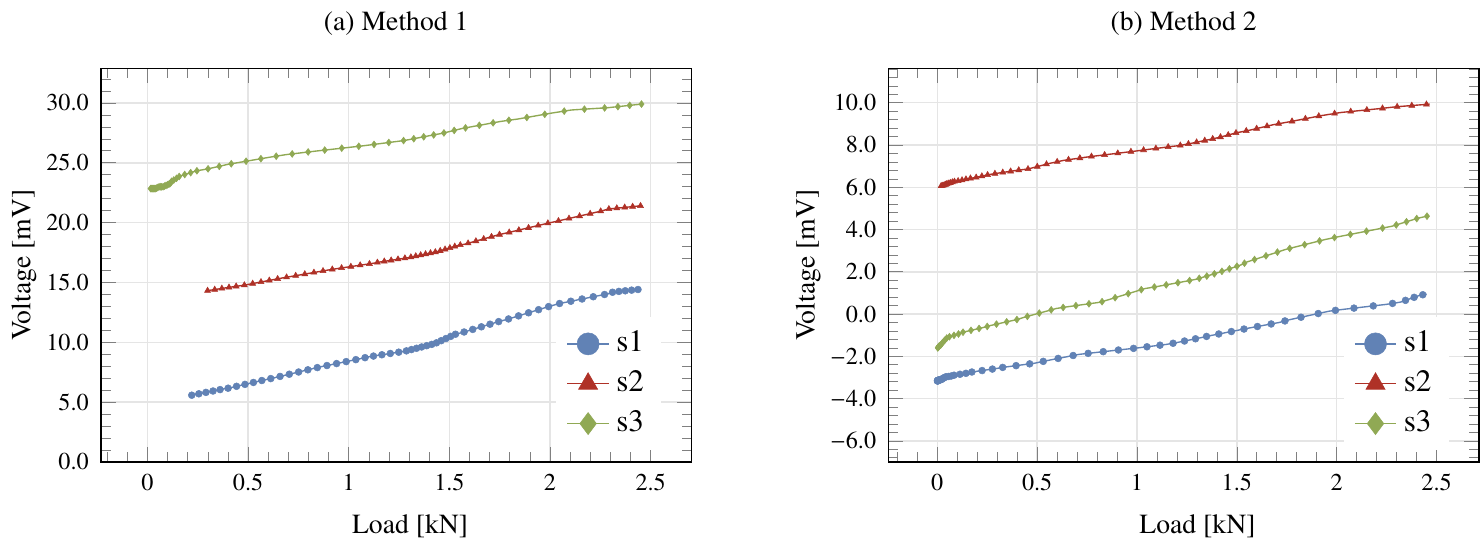}
\caption{Piezoelectric voltage versus compression load for rGO-based cement composites with rGO dispersed by (a) ultrasonic cleaner (Method 1) and (b) ultrasonic tip set (Method 2).
}
\label{voltage_load}
\end{figure}

Following the mechanical load history described in Fig.~\ref{experimental_setup} (c), Figs.~\ref{voltage_load} (a) and (b) report the output voltages versus the applied compression for dispersion Methods 1 and 2, respectively. It is first noted from these results that specimens fabricated by Method 1 (Fig.~\ref{voltage_load} (a)) generated the highest voltages, around three times larger than those generated by samples fabricated following Method 2 (Fig.~\ref{voltage_load} (b)). This is ascribed to the poorer filler dispersion obtained by Method 2 as previously reported in Section~\ref{fabrication}. The presence of rGO aggregates represents microstructural defects hampering the transportation of mobile ions in the composite, which affects the effective piezoelectric response. This effect is possibly compounded by the deterioration of the composite's effective mechanical properties as previously reported by Kiamahalleh \textit{et al.}~\cite{VALIZADEHKIAMAHALLEH2020118832}, who highlighted that rGO aggregates compromises the compatibility between the fillers and the host structure (cement). As a result, the filler-matrix load transfer capability is reduced, presumably affecting the piezoelectric properties of the composite as well. It is interesting to note in Fig.~\ref{voltage_load} that, despite the specimens were electrically discharged before testing, there exist some initial voltages indicating the presence of transferences of electric charges between the specimens and the environment. Therefore, we will concentrate on the linear voltage change with respect to the mechanical load. On this basis, the voltage piezoelectric coefficients $g_{33}$ have been extracted from the slopes of these curves as previously reported in Section~\ref{piezo_char}, and the estimated coefficients are collected in Table~\ref{piezo_factors}. 

\begin{figure}[ht]
    \centering
    \includegraphics[scale=1.1]{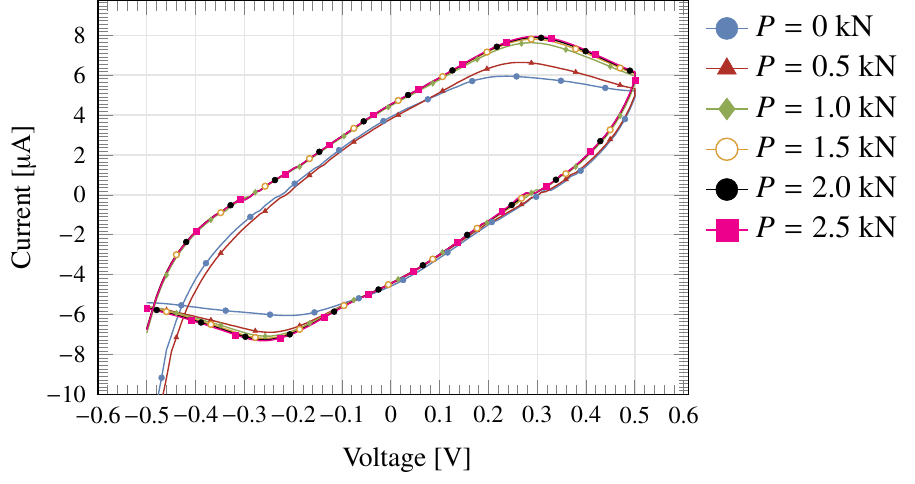}
    \caption{Cyclic voltammograms (scan rate: 25 mV/s) from rGO/cement specimen s1-M1 under increasing compression loads $P$.}
    \label{voltamp_p39}
\end{figure}

With the aim of computing the piezoelectric charge coefficients $d_{33}$ according to Eq.~(\ref{eq:d33}), cyclic voltammetry was conducted for all the different load steps in the loading sequence depicted in Fig.~\ref{experimental_setup} (c). As an illustrative example, the voltammograms obtained for s1-M1 are shown in Fig.~\ref{voltamp_p39}. In this figure, the sudden rises in the electric current following the lowest peaks in the voltammograms ($\approx$-0.26 V) indicate the presence of reduction-oxidation (redox) reactions through which rGO particles are reduced on the electrodes. On the other hand, the highest electric current peaks are related to the graphene oxidation process~\cite{KAUPPILA201384}, which occurs at $\approx$0.25 V. This is conceivably explained by the high content of oxygen in the rGO groups, which accelerates the redox process at the cement-rGO-electrode interfaces. Furthermore, when the specimen is subjected to increasing compressive loads, the voltage difference between the current peaks experiences slight increases. Specifically, the output voltage related to the electric current peak stemming from the graphene oxidation raises from 0.4 V (unloaded) to 0.51 V (compression load of 2.5 kN). This fact evidences that the capacitive component grows along with the strain-induced increment of the electrical conductivity of the rGO/cement composite.

\begin{figure}[ht]
\centering
\includegraphics[scale=1.0]{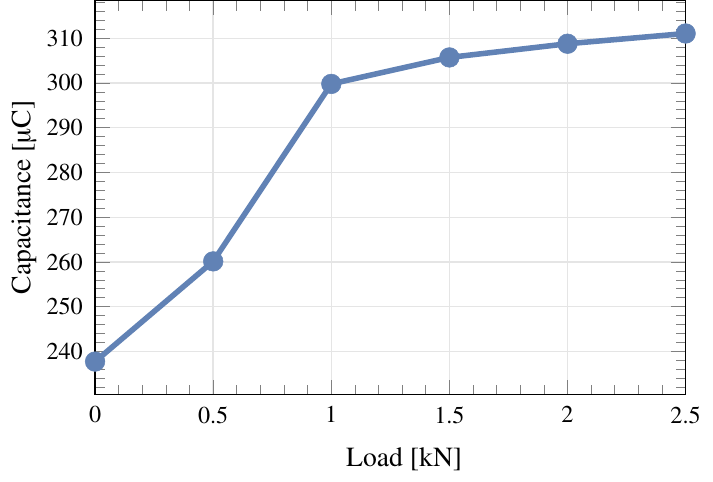}
\caption{Capacitance versus compression load for rGO/cement specimen s1-M1 obtained through cyclic voltammetry.}
\label{C_p39}
\end{figure}

The capacitances of the specimens were computed as functions of the exerted mechanical loads as previously indicated in Section~\ref{piezo_char}. To do so, the area under the voltammograms was calculated for every compression step according to Eq.~(\ref{cap}). The capacitance versus compression curve obtained from the voltammograms in Fig.~\ref{voltamp_p39} is shown as an illustrative example in Fig.~\ref{C_p39}. Note in this figure that the capacitance exhibits a clear non-linear behaviour with respect to the applied mechanical load. Specifically, the capacitance experiences considerable increases from 240 {\textmu}F until 300 {\textmu}F for mechanical loads raising from 0 kN to 1 kN. Nonetheless, the capacitance achieves a saturation zone for compression loads above 1 kN. In this region, the capacitance tends to converge to a stable value ($\approx$306 {\textmu}F), which indicates a limit in the creation of new electric dipoles responsible for the polarization properties of the rGO/cement composite. 

The exact same procedure was applied to all the manufactured specimens, and the resulting piezoelectric charge coefficients $d_{33}$ are reported in Table~\ref{piezo_factors}. For comparison purposes, the piezoelectric coefficients reported in references~\cite{AlQaralleh2022,Jaitanong2014, Gong2010} are also included in this table. Firstly, it is noted that the piezoelectric coefficients $g_{33}$ and $d_{33}$ are considerably larger for the specimens fabricated by Method 1 compared to those prepared with Method 2. In particular, the $d_{33}$ coefficient for Method 1 amounts to $1122.28$ $\pm246.67$ pC/N, a value about 2 times the coefficient obtained by Method 1, and 47 times larger that those reported in reference~\cite{Jaitanong2014} for PZT/cement composites. Such a high value indicates the relative tendency of the developed rGO/cement composites to hold the electric polarization state for a long period of time. On the other hand, the voltage piezoelectric coefficients $g_{33}$ are considerably lower than those reported for cement composites doped with piezoelectric ceramics in references~\cite{Jaitanong2014, Gong2010}. Nevertheless, note that the obtained coefficients $g_{33}$ are only slightly lower than the one reported by Al-Qaralleh~\cite{AlQaralleh2022} for cement paste polarized while curing under an external voltage of 5 V. 

It is important to mention that previous research has reported the potential effects of working environmental conditions on the piezoresistive properties of SSCCs (e.g. refer to~\cite{han2014self}). Specifically, as mentioned in the introduction, the work by Dong \textit{et al}.~\cite{DONG2019107488} highlighted the need to select an adequate water content to achieve optimal sensitivity in CB/cement sensors. However, the analysis of such effects upon the intrinsic piezoelectric properties of SSCCs has yet to be fully addressed, most works in the literature focusing on piezoceramic-cement composites. It is worth noting the work by Wittinanon and co-authors~\cite{wittinanon2023effect}, who reported on the effects of the w/c ratio on the piezoelectric properties of BZT/cement composites. Their results demonstrated that the piezoelectric coefficients increase with the w/c ratio, achieving maximum sensitivities of $d_{33}=16.5$ pC/N and $g_{33} = 13.1$E-3 Vm/N for the maximum ratio investigated of w/c=0.10. Future research efforts should focus on analysing the influence of temperature, humidity, and rheological properties of rGO/cement piezoelectric composites to devise effective strategies to compensate their influence on the output of the sensors.

\begin{table}[H]	
\setlength{\tabcolsep}{3pt} 
\newcommand\Tstrut{\rule{0pt}{0,3cm}}         
\newcommand\Bstrut{\rule[-0.15cm]{0pt}{0pt}}   
\footnotesize					
\caption{Piezoelectric voltage/charge coefficients, $g_{33}$ and $d_{33}$, and generated energy of rGO/cement samples manufactured following dispersion Methods 1 and 2. (Tolerances represent the standard deviation values obtained through ANOVA analysis of the three samples manufactured per dispersion method).}
\vspace{0.1cm}
\centering							
\begin{tabular}{cccc}	
\hline					    
Dispersion method & $g_{33}$           &   $d_{33}$ &  Generated energy density\Tstrut\Bstrut\\
                  &  [$10^{-5}$ mVm/N] &    [pC/N]  & [nJ/$\mathrm{cm^3}$] \Tstrut\Bstrut\\
\hline
         Method 1 & 13.02 $\mathrm{\pm}$0.66 & 1122.28 $\mathrm{\pm}$246.67 & 38.65 $\mathrm{\pm}$10.42 \Tstrut\Bstrut\\
         Method 2 & 5.02 $\mathrm{\pm}$1.18 & 593.11 $\mathrm{\pm}$147.42 & 10.30 $\mathrm{\pm}$2.61  \Tstrut\Bstrut\\
         Cement paste & 3.70~\cite{AlQaralleh2022} & - & -  \Tstrut\Bstrut\\
         Polarized cement (5 V) & 12.09~\cite{AlQaralleh2022} & - & -  \Tstrut\Bstrut\\
         PVDF/PZT/cement & 2570~\cite{Jaitanong2014} & 24~\cite{Jaitanong2014} & -  \Tstrut\Bstrut\\
         CNTs/PZT/rGO/cement & 6000~\cite{Gong2010} & - & -  \Tstrut\Bstrut\\
\hline
\end{tabular}			
\label{piezo_factors}							
\end{table}

\subsection{Electromechanical response under quasi-static compressive loading}\label{Sect22}

\begin{figure}[ht]
\centering
\includegraphics[scale=1.0]{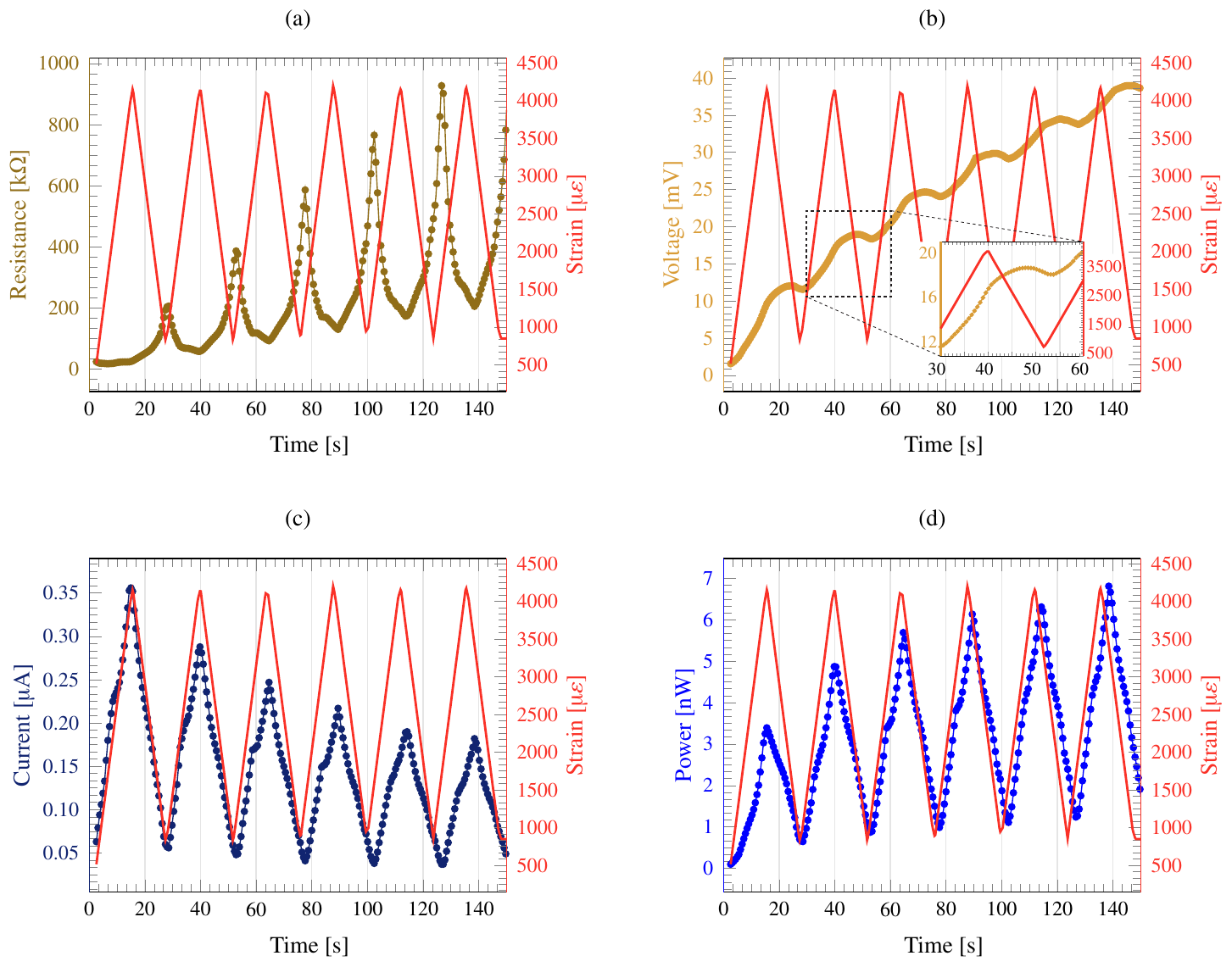}
\caption{Electrical resistance (a), piezoelectric voltage (b), electrical current (c) and generated power (d) time series measured from specimen s1-M1 and subjected to triangular cyclic loading (frequency of 0.05 Hz).}
\label{power_resistance}
\end{figure}

This section reports the electromechanical characterization tests conducted under quasi-static cyclic mechanical loading as described in Fig.~\ref{experimental_setup} (d); simultaneously, the piezo-voltage and generated electric current were measured by the potentiostat. As an illustrative example, the time series of electrical resistance, output voltage, current and power obtained for specimen s1-M1 are reported Fig.~\ref{power_resistance}. Note that, since no external power is supplied to the samples, the electric resistance is determined as the piezoelectric voltage divided by the generated electric current. In order to highlight the influence of the loading condition, the time series of mechanical strains are also included in this figure. It is noted in Fig.~\ref{power_resistance} (a) that the effective electrical resistance develops inversely proportional to the applied mechanical load, which agrees with the commonly observed piezoresistive behaviour in SSCCs. In this case, this can be conceivably explained by the coupled interaction of piezoresistivity and piezoelectricity. Note that piezoelectricity induces the appearance of electric currents, which are stored in the material due to capacitive effects and, in turn, cause increases in the apparent electrical resistance. Additionally, note that the electrical resistance exhibits increasing peaks as the loading cycles progress. Such increases in the electric resistance are not ascribed to polarization phenomena as typically observed in cementitious materials under external power supply. Instead, this electric resistance can be understood as the Thevenin resistance of the piezoelectric system. It is also interesting to note the generated voltage in Fig.~\ref{power_resistance} (b) exhibits a clear rising trend with some oscillating cycles corresponding to the loading/unloading peaks. A closer inspection of the oscillations in the zoom insert in Fig.~\ref{power_resistance} (b) reveals that, as the compression rises, the generated voltage increases almost linearly. Once the peak is achieved and the compression starts dropping, the piezoelectric voltage decreases non-linearly with a considerably slower rate. These decreases are possibly due to the strain-induced reduction in the amount of free electrical charge susceptible to be polarized into the material. Likewise, the available electrical charge to move across the electrodes also declines as it is consumed to produce either oxidation or reduction between the cooper electrodes and the composite. 

On the other hand, it is interesting to note in Fig.~\ref{power_resistance} (c) that the generated current exhibits a certain direct proportionality with the mechanical strain. This evidence contrasts with the classical theory of linear piezoelectricity, which describes the generated electrical current as directly proportional to the time derivative of the applied mechanical strain as reported above in Eq.~(\ref{piezoelec1}). Furthermore, note that the electrical current exhibits an almost exponential decrease in time, which denotes the existence of electrical damping due to inductance effects. Instead, the time series of electrical power exhibits higher linearity with the applied strain. Specifically, note in Fig.~\ref{power_resistance} (a) that, for each cycle of mechanical loading, the peaks in the time series of the electrical power increase until reaching an almost stationary value. After this, the generated electrical power exhibits clear linearity with respect to the applied strain. In this particular curve, these peaks correspond to a maximum piezoelectric-generated power of 6.8 nW (power density of 0.48 nW/$\mathrm{cm^3}$). Note that this value is substantially lower than the power density of 14 {\textmu}W/cm$^3$ reported by Kumar and co-authors~\cite{KUMAR2017174} for cementitious composites doped with 1wt\% PVDF and 1wt\% rGO. Such a considerable difference is conceivably due to the contribution of PVDF and the large concentration of rGO used in~\cite{KUMAR2017174}, which was four times higher than the content in this work. Nonetheless, although the limited power density of the developed composites may limit their use in energy harvesting applications, the high signal-to-noise ratios in the results in Fig.~\ref{power_resistance}, combined with the low power consumption of the utilized potentiostat, pave the way for the development of ultra-low power-consumption sensors. This allows categorizing this class of sensors as self-powered. Finally, regarding the comparison between the samples manufactured following dispersion Methods 1 and 2, the energy density values were calculated for all the samples as reported in Table~\ref{piezo_factors}. It can be concluded from these results that specimens prepared following dispersion Method 1 exhibit an energy production capability 73\% higher than those produced by Method 2.

\begin{figure}[ht]
\centering
\includegraphics[scale=1.0]{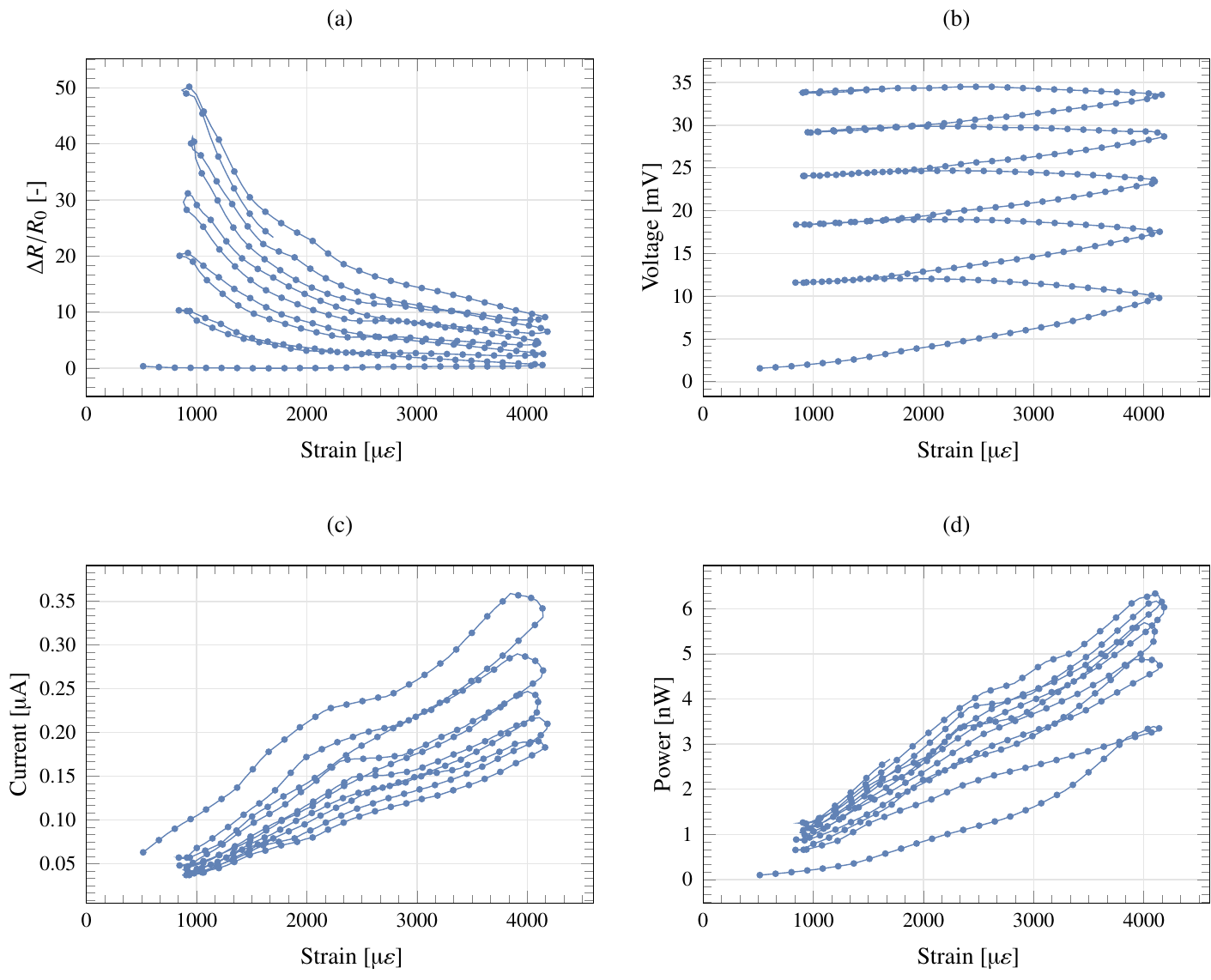}
\caption{Fractional change in resistance (a), piezoelectric voltage (b), electric current (c) and generated power (d) versus mechanical strain curves measured from specimen s1-M1 and subjected to triangular cyclic loading (frequency of 0.05 Hz).}
\label{sensor}
\end{figure}

With the aim of gaining some insight into the applicability of the developed rGO/cement composites for strain sensing applications, the electrical response of sample s1-M1 versus the applied mechanical strain is depicted in Fig.~\ref{sensor}. In particular, the response in terms of relative variation of electrical resistance, capacitance, current, voltage, and electric power are shown in Figs.~\ref{sensor} (a) to (d), respectively. It is clear in Fig.~\ref{sensor} (a) that, likewise other SSCCs (see e.g.~\cite{DaAlessandro2014}), the application of increasing compression loads leads to decreases in the effective electrical resistance of the material. Nonetheless, the fractional change in terms of electrical resistance exhibits a distinct non-linear behaviour compared to previous results in the literature. Note in Fig.~\ref{sensor} (a) that non-linearities concentrate for small strains ($<2500$\textmu$\varepsilon$), while the fractional change of electrical resistance tends to exhibit a more linear behaviour for medium to moderate strain levels ($>2500$\textmu$\varepsilon$). Different trends have been previously observed in the literature for cement- and polymer-based smart composites. For instance, Fu and Chung~\cite{fu1997effect} reported the opposite behaviour for cement mortar samples doped with carbon fibers under DC power supplies. Specifically, their results evidenced the appearance of high linearity for low mechanical strains ($\approx 0-1\text{\textperthousand}$), followed by increasing non-linear effects for larger strain values. In this work (no power supply is applied to the samples), instead, non-linearities are ascribed to the combined effect of piezoelectric and piezocapacitive branches as further analysed hereafter in Section~\ref{Sect23}. From a physical standpoint, piezoresistive effects interact with the strain-induced increases in the number of electric dipoles affecting the piezoelectric response, which conceivably gain relevance at low strain values. Specifically, the electric dipoles constrain the mobility of electrons through the piezoresistive branch so affecting the linear behaviour of the fractional change in electrical resistance. Furthermore, it is evident in Fig.~\ref{sensor} (a) the presence of certain hysteresis effects. In particular, while the fractional change in resistance raises until 1000\% after each discharge cycle at the minimum strain value, it only increases 2\% on average after the loading cycle (maximum strain value). Regarding the correlation between the output voltage and strain in Fig.~\ref{sensor} (b), it can be seen that the piezoelectric voltage increases almost monotonically with a rate of around 6 mV/cycle. Finally, let us focus on the correlation between the applied mechanical strain and the output current and generated power in Figs.~\ref{sensor} (c) and (d), respectively. It is evident in these figures the existence of certain hysteresis cycles, which are comparatively larger in terms of output current. In general, these magnitudes exhibit a considerably high level of linearity with respect to the applied strain, with Pearson's correlation coefficients $\mathrm{R^2}$ of 0.96 and 0.89 in terms of electric power and current, respectively. The presented results suggest the use of the generated electric power as a convenient magnitude for sensing applications in SHM.

\subsection{Lumped circuit model}\label{Sect23}

This section presents the experimental validation of the equivalent lumped circuit previously introduced in Section~\ref{Sect1}. In these analyses, the main parameters of the model $\left\{R_m, L, C_m, R_{ct}, \lambda_{R_m}, \lambda_{C_m}\right\}$ are estimated through inverse calibration using the experimentally assessed piezoelectric voltage curves presented hitherto. Specifically, the model calibration is set up as a minimization problem with an objective function or cost function involving the mean squared errors between the model predictions and the experimental data. The optimization problem is solved using the gradient-descent optimization algorithm implemented in Matlab Simulink environment. With the aim of minimizing ill-conditioning in the inverse problem and obtaining physically meaningful solutions, the model parameters are firstly bounded to physically realistic intervals. In particular, gauge factors $\lambda_{R_m}$ and $\lambda_{C_m}$ are constrained to the intervals $(0, \infty)$ and $(-\infty, 0)$, respectively. Regarding the intervals of the passive components, resistances were constrained to $\mathrm{(R_m > R_{ct})}$, while the capacitance $\mathrm{(C_m)}$ and the inductance $\mathrm{(L)}$ we forced to have positive values. In these analyses, given the previously reported suspicion on the applicability of the linear piezoelectricity theory, the electric current $i(t)$ in Eq.~(\ref{edofinal}) is directly taken from the experimental data. It is important to remark that good fittings were obtained by defining the piezoelectric current as linearly proportional to the applied mechanical strain. Nevertheless, the verification of the physical significance of such a relationship would deserve further investigation in a separate contribution, so no specific functional relationship between the piezoelectric current and the applied strain was imposed herein. The resulting model parameters for the samples fabricated following dispersion Methods 1 and 2 are reported in Table~\ref{model_parameters}. Overall, it is noted that no significant differences ($\approx 2-40\%$) were found between the values of $\lambda_{R_m}$, $\lambda_{C_m}$, and $C_m$ for dispersion Methods 1 and 2, while largest differences concentrate in the values of $L$, $R_m$, and $R_{ct}$. Note that the gauge factors in terms of electrical resistance $\lambda_{R_m}$ are considerably low, although in the same order of magnitude as those reported by Guo \textit{et al}.~\cite{guo2021study} for GO/cement composites (0.05-0.15 wt\%). Such low values are ascribed to the low rGO concentration used in this work, thereby the contribution of conductive networking is presumably limited and so the contribution of piezoresistive effects. It is also interesting to note in this table that the largest values of the cement matrix resistance $R_m$ are found for samples fabricated following dispersion Method 1 (27\% higher that those fabricated using Method 2), even though the dispersion of these samples was significantly better as previously shown in Fig.~\ref{ps}. This can be attributed to the low filler content adopted in this work, which, being well below the percolation threshold, causes the conductive networking mechanism to become ineffective. Therefore, the appearance of filler agglomerates may be more effective to increase the electrical conductivity as filler clusters favour the transfer of electric charges through quantum tunnelling effects. Finally, note in Table~\ref{model_parameters} that low inductance values $L$ in the order of $10^{-7}$ H are obtained for both methods. These values are considerably lower that those reported in other works in the literature on cement-based piezoelectric ceramic composites (see e.g.~\cite{XING20082456}). Nonetheless, the definition of the piezoelectric current directly from the experiments may limit the effective contribution of $L$. Thereby future further investigations are needed to evaluate the relationship between the piezoelectric current and the mechanical strain and, subsequently, the actual role of $L$ in reproducing electrical damping phenomena. 



\begin{table}[h]	
\setlength{\tabcolsep}{3pt} 
\newcommand\Tstrut{\rule{0pt}{0,3cm}}         
\newcommand\Bstrut{\rule[-0.15cm]{0pt}{0pt}}   
\footnotesize					
\caption{Fitted parameters of the lumped circuit model from dispersion Methods 1 and 2 (Tolerances represent the standard deviation values obtained through ANOVA analysis of the three samples manufactured per dispersion method).}
\vspace{0.1cm}
\centering							
\begin{tabular}{ccccccc}	
\hline					    
Dispersion & $L$ & $R_m$ &  $\lambda_{R_m} $ & $C_m $ & $\lambda_{C_m}$ & $R_{ct}$ \Tstrut\Bstrut\\
method     &  [{\textmu}H] & [$\mathrm{k\Omega}$] & [-] & [{\textmu}F] & [-] & [$\mathrm{k\Omega}$] [$\mathrm{10^{-2}}$] \Bstrut\\
\hline
Method 1 & 0.15 $\mathrm{\pm}$ 0.00 & 2899.10 $\mathrm{\pm}$ 205.89 & 1.59 $\mathrm{\pm}$ 0.31 & 307.93 $\mathrm{\pm}$ 80.49 & -4.96 $\mathrm{\pm}$ 2.17 & 10.97 $\mathrm{\pm}$ 2.08\Tstrut\\
Method 2 & 0.07 $\mathrm{\pm}$ 0.02 & 2116 $\mathrm{\pm}$ 67.85 & 2.21 $\mathrm{\pm}$ 0.40 & 416.19 $\mathrm{\pm}$ 36.83 & -5.09 $\mathrm{\pm}$ 1.48  & 6.10 $\mathrm{\pm}$ 1.36\Bstrut\\
\hline
\end{tabular}			
\label{model_parameters}							
\end{table} 

\begin{figure}[ht]
\centering
\includegraphics[scale=1.0]{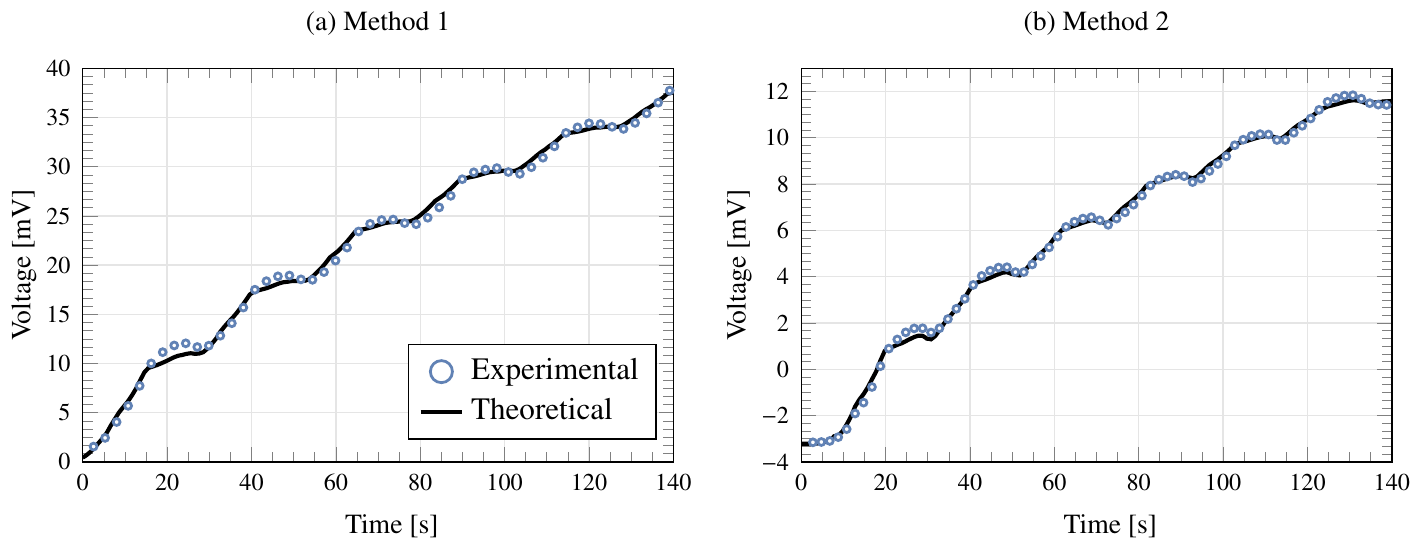}
\caption{Comparison between experimental and analytical predictions of piezoelectric voltage time series. Samples s1 fabricated by dispersion Methods 1 (a) and 2 (b).}
\label{model_figs}
\end{figure}

\begin{figure}[ht]
\centering
\includegraphics[scale=1.0]{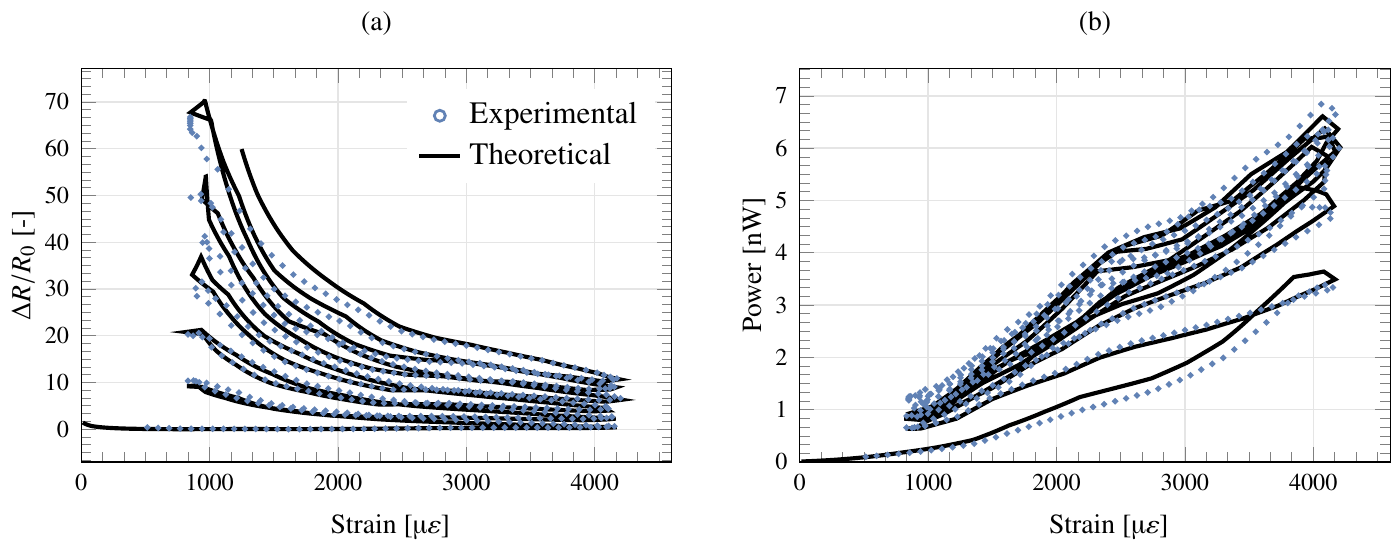}
\caption{Fractional change in resistance (a) and electric power (d) versus mechanical strain under triangular cyclic loading (frequency of 0.05 Hz). (Sample s1-M1).}
\label{model_figs_b}
\end{figure}

Figures~\ref{model_figs} (a) and (b) show the comparison between the experimental data and the predictions of the fitted circuital model in terms of output voltage for samples s1-M1 and s1-M2, respectively. Overall, it can be concluded from Fig.~\ref{model_figs} that the proposed circuital model can accurately represent the experimental data, including the two different regimes in the voltage curves previously discussed in Section~\ref{Sect22}. The model shows that the charge transfer resistance $R_{ct}$ dominates the piezoelectric voltage as it is evident from Eq.~(\ref{ouputvoltage}). Indeed, to replicate the larger voltage levels outputted by samples fabricated following dispersion Method 1 (in Fig.~\ref{model_figs} (a) the output voltage for s1-M1 is around four times the output voltage measured in s1-M2), the model calibration determined on average a $R_{ct}$ value 44\% (10.97 $\mathrm{k\Omega}$) larger than the one for dispersion Method 2 (6.10 $\mathrm{k\Omega}$). Similarly close fittings were also obtained in terms of electrical resistance and electric power as shown in Figs.~\ref{model_figs_b} (a) and (b), respectively. It is interesting to note in Fig.~\ref{model_figs_b} (a) that the circuital model can correctly describe the strong non-linear behaviour in the electrical resistance even though the linear piezoresistivity relation from Eq.~(\ref{eq:FCR}) is adopted. Note that the reported electrical current represents the Thevenin resistance or effective resistance interconnected in parallel with the piezoelectric current source $i(t)$. Therefore, these results strengthen the previous discussion on the origin of such non-linearities from the coupling between the piezoelectric/piezocapacitive/piezoresistive effects instead of a non-linear piezoresistive behaviour. Overall, it can be concluded that the presented lumped circuit model can accurately reproduce the output voltage, electrical resistance and generated power of rGO/cement composites, including the most significant non-linearities observed during the experimental campaign.


\section{Concluding remarks}\label{Sect3}

This work has presented the preparation and characterization of highly piezoelectric rGO-based SSCCs for self-powered strain monitoring applications. To this aim, a novel comprehensive methodology involving cyclic voltammetry and quasi-static compression testing has been proposed to characterize the piezoelectric coupling coefficients. The developed methodology has been applied to laboratory samples manufactured following two different filler dispersion methods, namely through ultrasonic cleaner and ultrasonic tip. Furthermore, a novel lumped circuital model has been proposed to replicate the electromechanical response of rGO/cement composites under cyclic compressive loads. The circuital model combines a piezoelectric current source in parallel with a RL series circuit accounting for piezocapacitive and piezoresistive effects. The key findings of this work can be summarized as follows:

\begin{itemize}
    \item The presented results and discussion have evidenced the role of the dispersion methodology of rGO/cement composites on the effective piezoelectric properties. It has been concluded that filler dispersion through ultrasonic cleaner produces more uniform rGO dispersions which, in turn, leads to higher piezoelectric voltage and energy generation capabilities.
    
    \item A novel methodology combining cyclic voltammetry measurements and mechanical testing has been developed to characterize the piezoelectric coefficients $d_{33}$ and $g_{33}$. The presented results have evidenced piezoelectric charge coefficients $d_{33}$ about 47 times larger than previously reported results in the literature for PZT-based cement composites.
    
    \item The presented experimental results have evidenced that, given the low filler content adopted in this work, the piezoelectric response plays a leading role in the strain-sensing capability of the developed composites. Specifically, the generated electrical power has been found as a candidate magnitude for conducting self-powered strain-sensing applications given its high linearity with the mechanical strain.
    
    \item The proposed circuital model has proved capable of reproducing the electromechanical response of the developed rGO/based composites in terms of piezoelectric voltage, electrical resistance, and electric power.
\end{itemize}

The proposed lumped circuit is envisioned to provide a useful tool for tuning the fabrication of rGO/based cement composites for self-powered SHM applications. Future research efforts shall address a further analysis of the correlation between piezoelectric current and externally applied mechanical strains. Furthermore, future investigation should cover the consideration of different rGO volume fractions covering the regimes below and above the percolation threshold, as well as different dynamic loading conditions to further validate the accuracy of the proposed circuital model.

\section*{Acknowledgements}
The authors would like to acknowledge the financial support from the \textit{Vicerrectoria de Investigación y Extensión} at the \textit{Universidad Industrial de Santander} for the grant project number 3711, entitled \textit{Dispositivo electrónico para caracterizar piezoeléctricamente cilindros de pasta de cemento}, which was contributed to the fabrication of a prototype for monitoring the sensors. Daniel A. Triana-Camacho want to give special thanks to scholarships program of \textit{MinCiencias} \textit{Programa de Becas de Excelencia Doctoral del Bicentenario - Corte 2, 2019}. E. Garc\'{\i}a-Mac\'{\i}as was supported by the Consejería de Transformación Económica, Conocimiento, Empresas y Universidades de la Junta de Andalucía (Spain) through the research project P18-RT-3128. The support and assistance of Prof. Felipe García from the University of Málaga (Spain) is also warmly acknowledged.

\bibliographystyle{elsarticle-num}

\bibliography{bibliography_clean}

\end{document}